# EXPRES: A Next Generation RV Spectrograph in the Search for Earth-like Worlds

C. Jurgenson*[a], D. Fischer[a], T.McCracken[a], D. Sawyer[a], A. Szymkowiak[a], A.B. Davis[a], G. Muller[b], F.Santoro[b]

[a]Yale University, Astronomy Dept., 52 Hillhouse Ave., New Haven, CT, USA 06511;
[b]ASTRO Electro-Mechanical Engineering, 1283 Carrizo St. NW, Los Lunas, NM USA 87031

## ABSTRACT

The EXtreme PREcision Spectrograph (EXPRES) is an optical fiber fed echelle instrument being designed and built at the Yale Exoplanet Laboratory to be installed on the 4.3-meter Discovery Channel Telescope operated by Lowell Observatory. The primary science driver for EXPRES is to detect Earth-like worlds around Sun-like stars. With this in mind, we are designing the spectrograph to have an instrumental precision of 15 cm/s so that the on-sky measurement precision (that includes modeling for RV noise from the star) can reach to better than 30 cm/s. This goal places challenging requirements on every aspect of the instrument development, including optomechanical design, environmental control, image stabilization, wavelength calibration, and data analysis. In this paper we describe our error budget, and instrument optomechanical design.

**Keywords:** precision radial velocity, white pupil spectrograph, double-scrambling, high resolution, laser frequency comb

## 1. INTRODUCTION: THE 100 EARTHS PROJECT

The design and construction of EXPRES is funded by the National Science Foundation Major Research Instrumentation program. The primary purpose will be to serve as the work-horse instrument for the 100 Earths Project. This project will search for planets that are similar in mass to our world, and that orbit their host stars at a similar distance where liquid water might flow in rivers and oceans. The NASA Kepler mission has been searching stars that are several hundred light years away and has demonstrated that Earth-sized planets are common. Armed with this important statistical information, we will take a census of the nearest neighboring stars to find terrestrial worlds. These planets will be the targets of intense searches for life outside the solar system.

### 1.1 The Path to Finding Earth Analogs

A radial velocity (RV) measurement precision of 10 cm/s is required in order to detect Earth analogs. To reach this precision, the 100 Earths Project will build upon the state-of-the-art[1] by advancing developments in six key areas: 1) instrument environmental stability, 2) stable light coupling, 3) high spectral resolution, line spread sampling, and signal to noise, 4) precision wavelength calibration, 5) removal of telluric contamination and stellar jitter, 6) near nightly observational cadence of target stars, and 7) new statistical treatment of stellar jitter. The spectrograph optical bench will be under vacuum in a climate controlled room on a vibration isolated slab. EXPRES will have a resolving power of 150,000, and < 0.01 Å per pixel sampling to improve spectral line modeling and detecting and decorrelating stellar activity signals. The use of a Menlo Systems laser frequency comb (LFC) for wavelength calibration will enable access to the majority of the instrument free spectral range (380 to 680 nm), leading to higher Signal to Noise Ratio (**SNR**) and more information about stellar activity. Success is not only coupled to the instrument design, but the implementation of robust statistical and modeling techniques.

### 1.2 Lowell Observatory – The Discovery Channel Telescope

Yale University is a partner with Lowell Observatory to use the newly commissioned 4.3-meter Discovery Channel Telescope (**DCT**)[2] for the 100 Earths Project. The spectrograph will be housed in a custom-built, environmentally stabilized room, and fed via fibers from the telescope interface. The EXPRES telescope interface will be permanently mounted on one of the five DCT instrument cube ports. Each port has its own deployable tertiary that will enable queue scheduling, therefore providing near-nightly observational cadence of target stars. Lowell Observatory operates the DCT in the Coconino National Forest, nearly 40 miles southeast of Flagstaff, Arizona. Near the community of Happy Jack, the observatory sits on a peak at 2360 meters elevation. The left panel in Figure 1 shows the DCT from the observing platform, while the right panel is outside the building.

*colby.jurgenson@yale.edu; phone 1 203 436-2848; fax 1 203 432-5048; http://exoplanets.astro.yale.edu

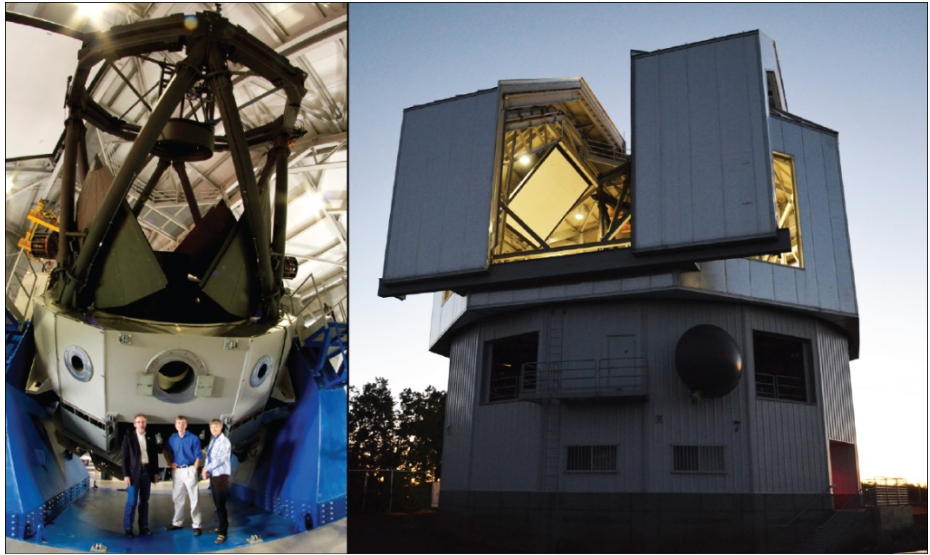

**Figure 1**. *Left*: the 4.3 – meter Discovery Channel Telescope operated by Lowell Observatory near Flagstaff, AZ. *Right*: the Discovery Channel enclosure building.

## 2. TOP LEVEL REQUIREMENTS

To achieve the science goals of the 100-Earths project, EXPRES must push the state of the art to produce spectra with unprecedented fidelity (high resolution, high SNR, stable PSF). The Doppler precision required to detect Earth-like exoplanets places very challenging requirements on the instrument stability, illumination uniformity, sensitivity, wavelength calibration, and data analysis techniques. EXPRES will have to build upon the technological advances made with other extremely stable instruments, such as the HARPS spectrograph[3,4] to push the limits of resolution and sensitivity, and pioneer new analysis techniques for distinguishing photospheric velocities from Keplerian velocities[1].

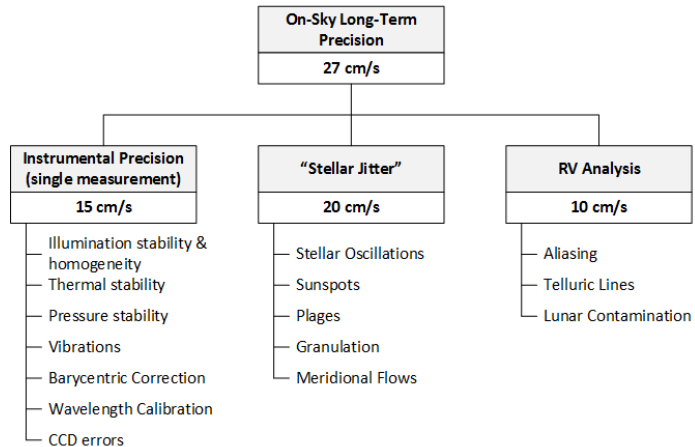

**Figure 2**. Top-Level error budget allocations and examples of terms that contribute to them.

The goal of EXPRES is to achieve the instrumental precision and spectral fidelity required to separate the wobbles in the star caused by planets from the Doppler noise originating from the photosphere of the star. **Figure 2** shows our top level radial velocity error budget allocations and examples of the terms that contribute to them. To achieve a long-term measurement precision of 27 cm/s, which will enable the detection of small rocky planets around late G and K type stars, our requirement for the instrumental precision is 15 cm/s and our goal is 10 cm/s. This instrumental precision drives the technical design of EXPRES, which is the focus of this paper.

In addition to the key requirements set by the instrumental RV precision, the error terms for stellar jitter and RV analysis also flow down to technical requirements for the instrument. The stellar jitter, which includes error terms that must be extracted and corrected in the spectra before attempting to extract Keplerian motions, drive key requirements for instrument resolution and efficiency (SNR). The error terms for RV analysis drive key requirements for instrument bandpass and observing strategies. These key and driving requirements are described in more detail in the following subsections.

## 2.1 Instrumental RV Precision

A detailed error budget is maintained for EXPRES to ensure that our requirement of 15 cm/s for instrumental RV precision is met. Our error budget has been developed using terms and allocations that have either been verified in our previous instruments or based on results published by other instrument teams[5]. Rather than reiterating the details of the specific error terms in our RV error budget, we will instead describe the key requirements that are derived from the error budget allocations.

Environmental stabilization is essential for achieving high RV precision. Changes in pressure and temperature lead to variations in the position of the spectrum on the CCD that could lead to large errors in the RV measurements. Like other stabilized spectrographs, EXPRES will be housed in a large vacuum chamber maintained at a pressure better than $10^{-3}$ Torr. In addition, the vacuum chamber will be located in a dedicated room that is thermally stabilized to +/- 0.5K and incorporates radiation shields and an insulating shell to passively maintain the spectrograph components to a +/- 1.0 mK stability.

There are many layers of vibration isolation required for EXPRES. The vacuum chamber will be located on the ground level of the DCT enclosure and will be mounted on an isolated concrete slab to decouple it from the enclosure foundation, which may transmit vibrations originating from dome rotation or wind buffeting. The vacuum chamber itself provides further vibration isolation as it sits on spring isolators. Inside the vacuum chamber, the optical bench is isolated from the vacuum vessel to dampen vibrations transmitted by sources on or near the structure, such as fans or motors. Finally, the CCD cryostat must decouple any vibrations produced by cryo-cooling from the CCD which is mounted on the optical bench.

Stable coupling of light is crucial to achieving high RV precision. This stable coupling starts in the front-end module where atmospheric dispersion compensation and fast tip-tilt control ensure stable illumination of the science fiber. Without an ADC, the photocenter of the image might not fall at the center of the fiber at all the wavelengths. This could change the measured spectral energy distribution, introducing time-correlated changes in the slopes of the echelle orders. Fast tip-tilt is necessary to correct for coherent image motion caused by the telescope and/or atmosphere and stabilize the image on the fiber. Any image motion on the fiber could result in a changing illumination pattern in the spectrograph and introduce instabilities in the line positions. Once the light is in the fiber, further stabilization is required using fiber scrambling methods that provide a scrambling gain of at least 5000[6]. Once propagated into the spectrograph, the light path requires careful placement of baffles and application of surface treatments to prevent stray light from introducing non-uniformities in the illumination.

To account for changes in radial velocity due to the Earth's rotation, EXPRES must incorporate an exposure meter to measure the photon-weighted mid-point for each observation to an accuracy of 0.25s. This accuracy is required to meet a budget allocation of 1 cm/s for exposures that are tens of minutes long[7]. Exposure times will be limited to 20-30 minutes to prevent the Barycentric motion from broadening the lines and degrading the RV precision.

The wavelength calibration of the spectrum is a key aspect to achieving our required RV precision. The laser frequency comb (LFC) provides the best solution because it produces a dense grid of stable, equally spaced, narrow lines of similar brightness. The LFC is currently available as a commercial turnkey system with a free spectral range (**FSR**) that is well matched to the resolution (R=150,000) and spectral sampling (4 pixels) of EXPRES to provide a calibration line every ten pixels or so. Our budget allocates a 3 cm/s calibration error to the LFC based on recent results achieved using the HARPS spectrograph[8], and assumes that the calibration will be effective in removing 90% of the calibratable errors[5].

For maximum flexibility, EXPRES is required to support two modes of LFC calibration: First, one can interleave science observations with the LFC calibration exposures by injecting the comb into the science fiber at the telescope focus. This has the advantage that the same light path and pixels used to obtain the science observation are being used for the wavelength calibration. The disadvantage is a hit to the duty cycle as this method will add a minute of calibration time at frequent intervals during the night. The second LFC calibration technique (implemented on HARPS and ESPRESSO) is to inject the calibration comb through a second fiber (about 10 pixel separation from the science fiber) called a "sim" or simultaneous fiber. The sim fiber is calibrated against an LFC comb through the science fiber at the beginning and end of the night and any drifts are taken out in software the next day. This has the advantage of improving the duty cycle of the observations and the disadvantage of calibration pixels that are different from those used for the science observations.

The EXPRES RV precision requirement places several constraints on the CCD detector system. Radial velocity measurements are extremely sensitive to charge transfer efficiency (**CTE**) effects, and any variations in CTE, such as dependence of CTE on the amount of carried charge, reduces the effectiveness of wavelength calibrations. Our budget allocation requires a worst case CTE of better than 99.9997% at any charge level, and our goal is to do even better. Fixed pattern noise, such as pixel size non-uniformity and pixel "stitching" errors can also degrade the precision of wavelength calibration as line features may move across different pixels, thus pixel-to-pixel uniformity is an important consideration. Finally, the mechanical stability of the CCD is extremely important as a lateral position shift of only 2 nm could introduce a 10 cm/s error.

## 2.2 Stellar Jitter

In order to detect the low amplitude radial velocity signals arising from gravitational reflex caused by Earthlike planets it is necessary to distinguish the photopheric velocities ("stellar jitter") from Keplerian velocities. Fortunately, stellar jitter has two important properties that can be exploited:

- it is not a persistent Keplerian signal; it waxes and wanes, it is not perfectly coherent, and it varies on timescales that are different from center of mass radial velocities.
- the underlying physical phenomena that spawn jitter have detailed spectroscopic, photometric, wavelength dependent, and polarization signatures that are, in principle, distinguishable from simple wavelength shifts due to Keplerian Doppler shifts.

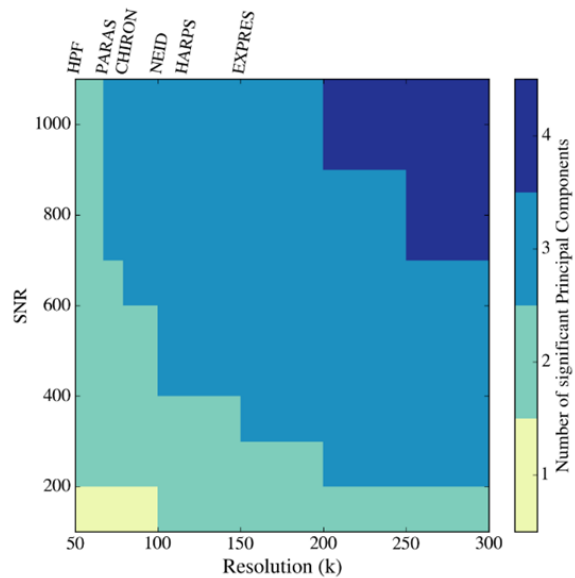

**Figure 3**. We carried out principal component analysis (PCA) of simulated spectra that were degraded over a range of spectral resolutions and SNR to check our recovery of PCs. The PC analysis is a test of the information content of the spectra as a function of instrument resolution and SNR and shows that at high resolution and high SNR, we can distinguish the effects from planets (orbital Doppler shifts) and photospheric velocities.

To help inform the design of EXPRES, our team has been exploring techniques for characterizing stellar jitter. With current instruments, all of the information in a spectrum is distilled into a single number: the radial velocity of the observation. This single number includes velocity signals from the photosphere as well as orbital velocity signals from prospective planets. Presently, astronomers try to decorrelate the spurious RV scatter using diagnostic information (the line bisector, the FWHM of the cross correlation function, or emission in spectral lines that form in the lower chromosphere such as Ca H&K II or H-alpha line core emission). This approach works reasonably well for detecting signals with amplitudes greater than 1 m/s, but to make a significant step toward 10 cm/s precision, we believe that the photospheric noise must be extracted from the hundreds of thousands of pixels that make up a spectrum, not from a single radial velocity number for each spectrum. Our approach applies techniques used by statisticians, such as dictionary learning with sparse representation, to extract an unperturbed picture from a noisy image.

We have carried out principal component analysis (**PCA**), a technique that is based in linear algebra that identifies orthogonal axes of maximum variability, on simulated time-series spectra. The PCA is a test of the information content of the spectra as a function of instrument resolution and SNR and through our simulations we find that with a high resolution and high SNR, the perturbing effects separate out into different eigenvectors, which confirms that we can distinguish the effects from planets (orbital Doppler shifts) and photospheric velocities[9].

Using the PCA approach, we degraded the resolution and SNR of the simulated spectra and analyzed them for the resulting information content. The number of recovered principal components, shown in Error! Reference source not found., captures the information content of our spectra. Photospheric signals are imprinted in spectra differently than Keplerian Doppler shifts (spectral line excitation potentials, wavelength dependence, temporal variations). For instrumental resolutions less than 100k and SNR < 200 (yellow region), we see that there is simply not enough information in the spectra to distinguish photospheric signals from orbital signals. At the resolution of HARPS, R=120k with SNR < 400 (green region), two principal components are recovered, which demonstrates a marginal improvement in the information content. With the requirements of EXPRES, R=150k at SNR > 400 (light blue region), we aim to move into a new regime of information content.

Achieving a high SNR with EXPRES was a driving factor during the design process. To maximize the throughput, our design incorporates a cross dispersing prism, instead of a VPH grating, and high performance AR coatings on all optical lens elements. The calculated throughput of the instrument and the system (instrument and telescope) is shown in Figure 4. In addition, we minimized the number of image slices (a trade for larger optical elements to maintain the resolution of R=150k) which minimized the length of the slit and allowed us to maximize the flux per pixel on the detector for a given spectral sampling of 4 pixels.

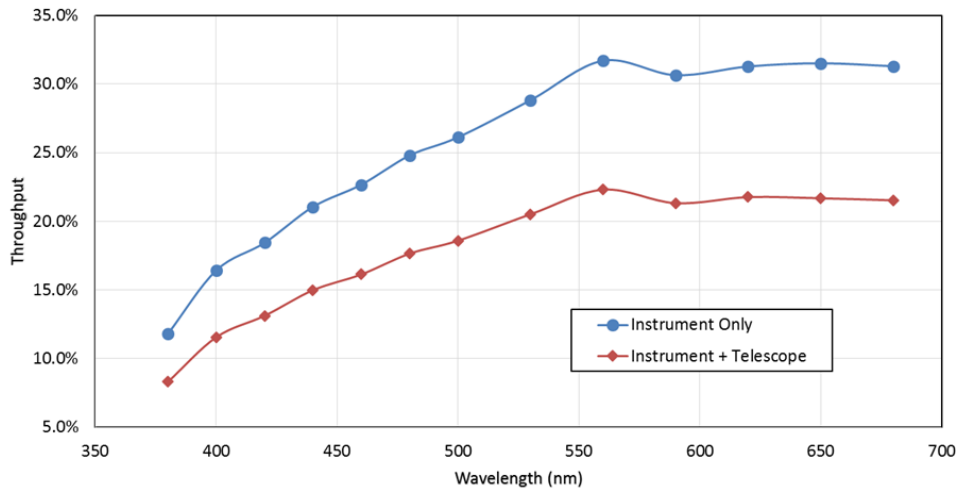

**Figure 4**. The calculated throughput of the EXPRES instrument and the total system when installed on the DCT telescope.

Exposure time durations must be limited to 20-30 minutes because the barycentric motion errors begin to smear out the lines for longer exposures. Thus the instrument efficiency must allow the desired SNR=400 to be achieved within the exposure time limit. An exposure time calculator (**ETC**) was developed to estimate the expected SNR performance as a function of stellar magnitude and exposure time when EXPRES is installed on the DCT. The ETC is comprehensive and includes realistic estimates for all factors that may impact the instrument efficiency, such as the average echelle efficiency (not blaze peak), a 60-meter science fiber used to feed the spectrograph from the front-end module mounted on the telescope, the instrument noise sources, the telescope factors (e.g. effective area, atmospheric extinction and telescope mirror losses) and a range of seeing conditions expected for the DCT site. **Figure 5** shows a plot of the estimated SNR as a function of exposure time for stellar magnitudes of $m_v=7$ and $m_v=10$.

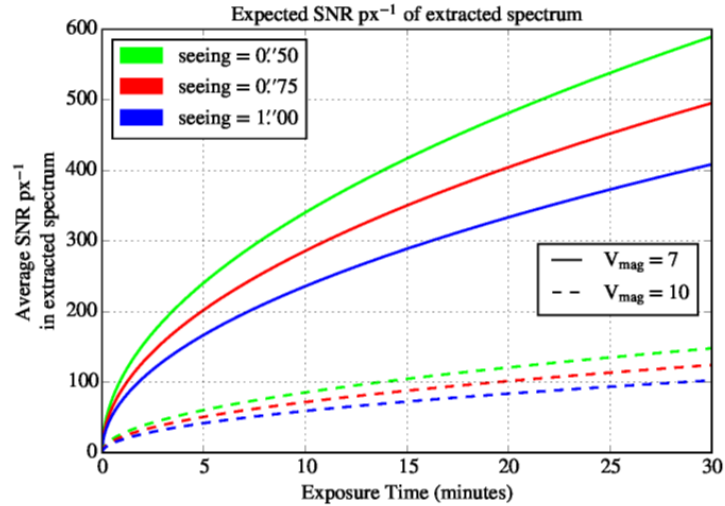

**Figure 5**. SNR per pixel in the extracted spectrum as a function of exposure time for various seeing conditions and apparent stellar magnitudes. These calculations include the estimated telescope mirror reflectivity and atmospheric extinction.

The ETC shows that, under good seeing conditions at the DCT site (0."9 corrected to 0."76 with fast tip-tilt), stars brighter than 7th magnitude will achieve the required SNR per pixel of 400 in a single exposure of about 20 minutes. EXPRES will be used to detect Earth-like planets at habitable zone distances orbiting late G or K dwarf stars. The late G and K dwarfs are better targets than solar type stars for the following reasons:

- the mass of the star is lower so the reflex velocity is larger,
- the habitable zone is closer to the star so the orbital periods are shorter; this proximity offers an additional boost to the reflex stellar velocity,
- the intrinsic stellar jitter (astrophysical noise from spots and flows) of late G and K dwarf stars is significantly lower than for solar type and more massive stars[10].

Under these conditions, EXPRES will make significant strides toward detecting 100 Earths. The reflex velocity of a 1 $M_{EARTH}$ planet orbiting late G and K dwarf stars at habitable zone distances is 15 - 20 cm/s, which is within the reach of the instrument, and there are hundreds of nearby late G and K dwarf stars at $m_V=7$ or brighter.

### 2.3 RV Analysis

The on-sky measurement precision of EXPRES depends on how well the analysis process can extract the Doppler velocities from the spectra. To improve the accuracy of the extraction process EXPRES includes requirements for a broad science bandpass, novel flat-field calibration features, methods for solar and Moon contamination, and observing strategies for high observation cadence.

EXPRES has a requirement for a science bandpass of 380nm to 680nm. This broad wavelength range provides significant Doppler information in every spectrum as it is rich in absorption lines from late G and K dwarf stars. The many lines available in the broad spectrum allow better statistics for analyzing the Keplerian motions. In addition, the blue limit of the bandpass is set to include Ca II H & K lines, which are good monitors of chromospheric activity, and provide further diagnostics for stellar jitter corrections.

Flat fielding is inherently problematic with spectrographs because a white light source projected through the science path is dispersed into the same spectral orders as the science spectra with the same intensity profile. Thus, in the wings of the orders the SNR of the flat-field reduces significantly and information contained there is lost. With EXPRES, we require the flat-field spectra to overfill the science orders so that effective flat-field reduction can be achieved all the way to the edge of the science orders. In addition, we require that the response of the white light source used for flat-fielding be tuned to inverse of the instrument response so that the resultant flat-field spectrum has a uniform intensity across the entire bandpass of the instrument. This will allow the SNR to be maximized at all wavelengths in a single flat-field exposure.

Contamination from telluric lines and lunar reflection are variable and can complicate the analysis process. Telluric absorption lines are imprinted in a stellar spectrum and shift across the spectral lines because of the barycentric velocity of the Earth. Telluric contamination perturbs the shape of the spectral line spread profile and must be properly treated for high precision radial velocity measurements. While stellar photospheric lines are broadened by pressure and rotation of the star, telluric lines are intrinsically narrow. At R=150,000, it will be much easier to distinguish telluric lines and to treat them by masking or fitting. We have developed an empirical code that uses PCA to identify and model out the unsaturated telluric lines. Saturated features can only be masked out. The lunar contamination adds a large variability in the line positions due to the motion of the Moon. With EXPRES, this will be addressed in the scheduling process where target selection will be prioritized based on proximity to the Moon. In addition, any observations within a certain distance of the Moon will be rejected by the reduction software.

It is very important to get many observations of exoplanet candidates on a frequent basis to eliminate aliasing effects and increase detection confidence. Thus, it is a requirement for EXPRES to be permanently mounted on the DCT, quickly ready for use during any night, and fully automated for efficient observations. An automated scheduling program, adapted from the CHIRON scheduler[11] will be used to optimize the target observations given the available telescope time. The operation of EXPRES will be highly automated to allow telescope operators or "service" observers to easily and efficiently observe scheduled targets, or to allow remote observers to operate the instrument.

## 3. INSTRUMENT OVERIVEW

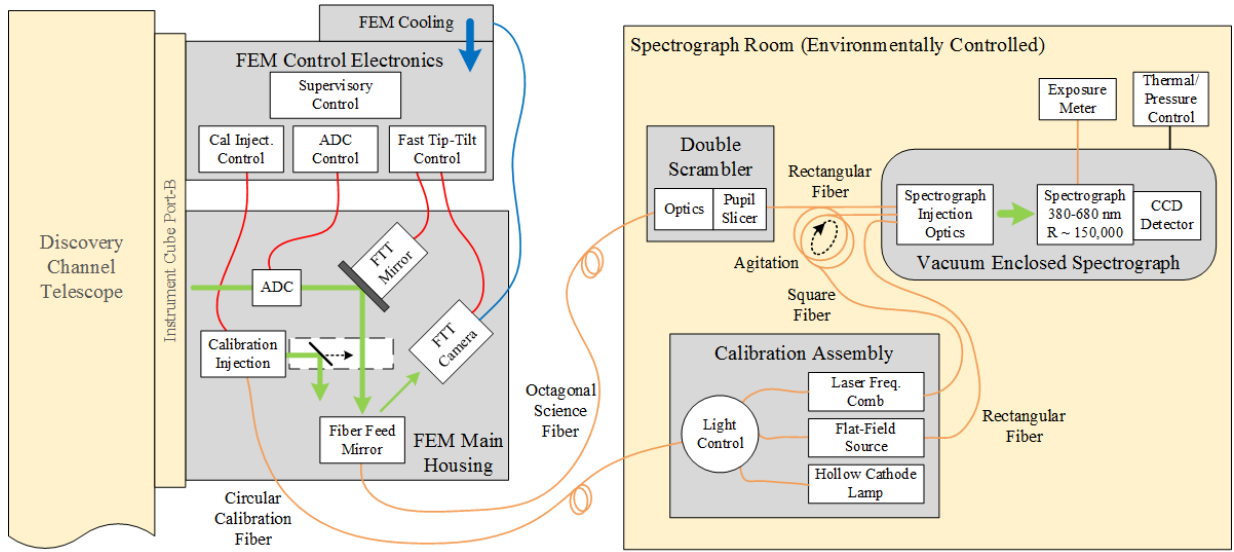

**Figure 6**. EXPRES has four distinct optomechanical subsystems. The front-end module interfaces to the telescope instrument cube. The calibration assembly, double-scrambler, and spectrograph all reside in an environmentally stabilized room on the ground floor of the observatory.

### 3.1 Instrument Architecture

The EXPRES instrument architecture is shown schematically in Figure 6. There are four distinct optomechanical subsystems that are connected with optical fibers of various sizes and shapes:

- Front-End Module (**FEM**): The FEM is attached directly to one of the telescopes instrument cube ports. Its primary functions are to correct for atmospheric dispersion, reimage the telescope beam onto the science fiber, stabilize the image with fast tip-tilt corrections, and to inject calibration light from the calibration assembly.

- Double-Scrambler/Pupil Slicer: Light from the telescope is transported to the spectrograph room via an octagonal fiber for near-field scrambling. Upon exiting the octagonal fiber it is sliced in the pupil plane, and reimaged onto a rectangular fiber. This not only serves as the second stage of scrambling, but achieves a resolving power of 150,000.

- Calibration Assembly: The calibration assembly enables the injection of three different light sources. A laser frequency comb (**LFC**), flat-fielding source, and thorium argon lamp. They can be injected into the science fiber out at the FEM, or directly into the spectrograph. A square fiber with LFC light is paired with the rectangular science fiber for simultaneous tracking of instrumental drifts. An extended flat fiber assembly can transport light into the spectrograph for two-dimensional flat-fielding without disturbing the science path.

- Spectrograph: The spectrograph vacuum enclosure is vibrationally and thermally isolated from the spectrograph room. There are no windows in the enclosure for light to pass through. The fibers either pass light through hermetically sealed connectors, or have a continuous feed (in the case of the rectangular science fiber). There is also a feed-through for a multi-channel exposure meter.

## 3.2 Optical Design

### 3.2.1 Front-End Module

**Figure 7** is the optical layout for the FEM. Light exiting the telescope comes to focus following a fold mirror at f/6.1. An off-axis parabola (**OAP**) collimates the beam, sending it through two pairs of prisms to correct for atmospheric dispersion, before forming an image of the DCT primary on the fast tip-tip (**FTT**) mirror. The FTT mirror then directs the light to a triplet that focuses the beam at f/3.5 onto an octagonal science fiber. It is 66µm edge-to-edge, and subtends 0.9 arcseconds on the sky. The fiber has been sized such that the central 84% of the radiation is injected into it. The outer halo and remainder of the 80 arcsecond field is reflected to the FTT camera.

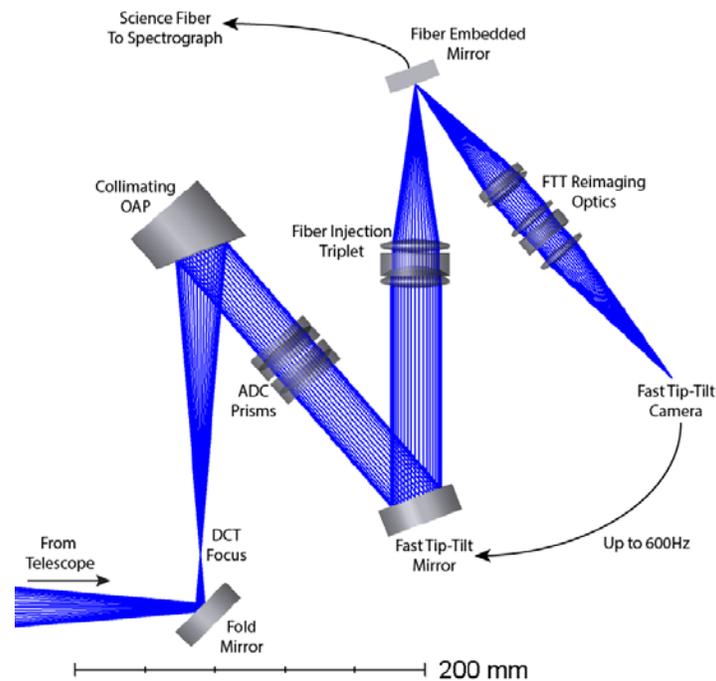

**Figure 7**. The FEM optical layout. Light from the telescope comes to focus following a fold mirror and is then collimated. It is directed through an ADC before FTT corrections are applied to ensure the starlight is stabilized on the fiber core.

The fiber itself is embedded into a block of aluminum whose face is diamond turned to serve as a mirror. In this way, the FTT camera is directly imaging the core of the fiber and feeding back to the FTT mirror to ensure that the image of the star is stabilized on the core. A doublet collimates the beam from the fiber mirror before it is reimaged at f/4.8 by a triplet. This yields a plate-scale of 0.16 arcseconds per pixel, spreading an image of the fiber core to just under six pixels. All lenses and ADC prisms are being fabricated and AR coated by the Nikon Corporation. The materials have high transmission in the blue region of the spectrum, and their AR coatings ($R_{avg}$ = 0.3%) are optimized for the individual materials and angles of incidence. Specially designed dielectric coatings (R = 99.5%) have been applied to the telescope tertiary, fold, and FTT mirrors to maximize reflectance across the band.

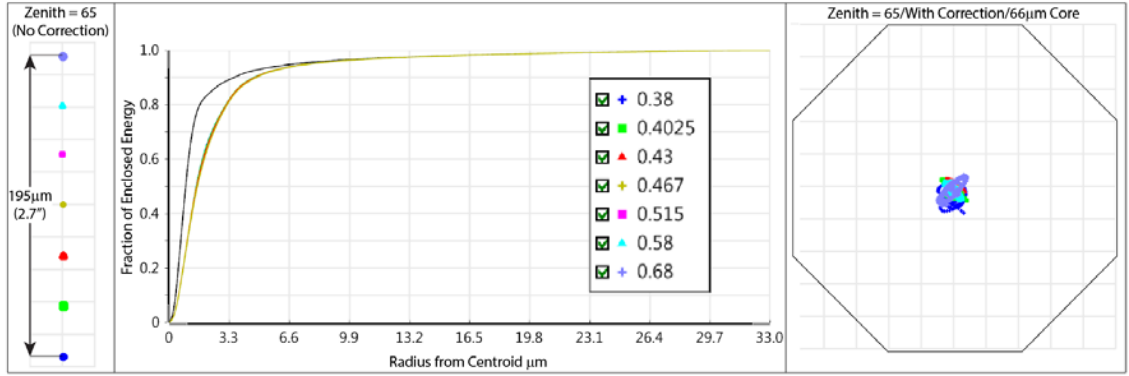

**Figure 8**. *Left*: Spot diagram showing the amount of uncorrected atmospheric dispersion for seven wavelengths between 380 and 680nm. *Center*: Diffraction encircled energy for the optimized system. *Right*: Spot diagram for the seven wavelengths shown to scale with the core of the octagonal fiber.

The left panel of **Figure 8** shows the resulting atmospheric dispersion at the maximum zenith angle of 65° in the focal plane of the science fiber. A total of seven wavelengths from 380 to 680nm (bottom to top spots in the panel) were modeled. Recall that the science fiber itself is only 66µm in diameter (0.9 arcseconds on sky). The fiber injection triplet was optimized for eleven different Zenith positions from 0 to 65°. The center and right panels of the figure show the diffraction ensquared energy and spot diagram for the worst case design performance (Zenith angle equal to 65°). The right panel also shows the scale of the 66µm octagonal fiber core relative to the spot sizes for the seven wavelengths. The residual error in the centroids for the seven wavelengths is less than 3µm.

### 3.2.2 Spectrograph

**Figure 9** shows the asymmetric white-pupil[12] optical layout of the spectrograph. Light is transported from the double-scrambler/pupil slicer (Section 4.2) via a 33x132µm rectangular fiber. It exits the fiber at f/3, and is converted to f/8.5 by cemented doublet. In addition to the rectangular science fiber, there is a 33x33µm square fiber that provides a simultaneous feed from the laser frequency comb. The doublet forms a virtual image of the rectangular and square fibers that is coincident with the focal point of the main collimator. This is an off-axis parabola (OAP) with a parent focal length of 1524mm, and a turning angle of 7.87°. Following the main collimator, a beam 180mm in diameter intercepts the R4, echelle grating with 31.6 grooves per mm to yield a resolving power of 150,000.

The echelle grating has a γ-tilt (about an axis perpendicular to the page) of 0.5° which slightly offsets the return beam to the main collimator from the input beam. This gives access to an intermediary focus, and allows the now dispersed beam to be directed to the cross-dispersing and camera optics. A cylindrical Mangin mirror intercepts the converging beam and directs it to the transfer collimator. The Mangin mirror is depicted in the left and middle panels of **Figure 10**. It is composed of NIGS7054 glass, has a concave cylindrical radius of curvature (**ROC**) on the front surface, and a convex cylindrical ROC on the rear surface. The rear surface is over coated silver so that the beam gets refracted by the first surface, reflects off of the rear, and then refracts back out the first surface. The purpose of the Mangin mirror is to remove the field curvature prior to the to the camera optics. Typically a cylindrical field flattening lens is placed prior to the detector focal plane, but the performance is not as good. The right panel of **Figure 10** displays the spot diagram at 680nm prior to ROC optimization (above), and after (below). The optimization is a simple spot radius reduction, varying the cylindrical radii and the material. A paraxial (ideal) lens serves in place of the camera optics so that it does not have to account for aberrations due to those elements. Almost a factor of 10 improvement in the RMS spot radius is observed.

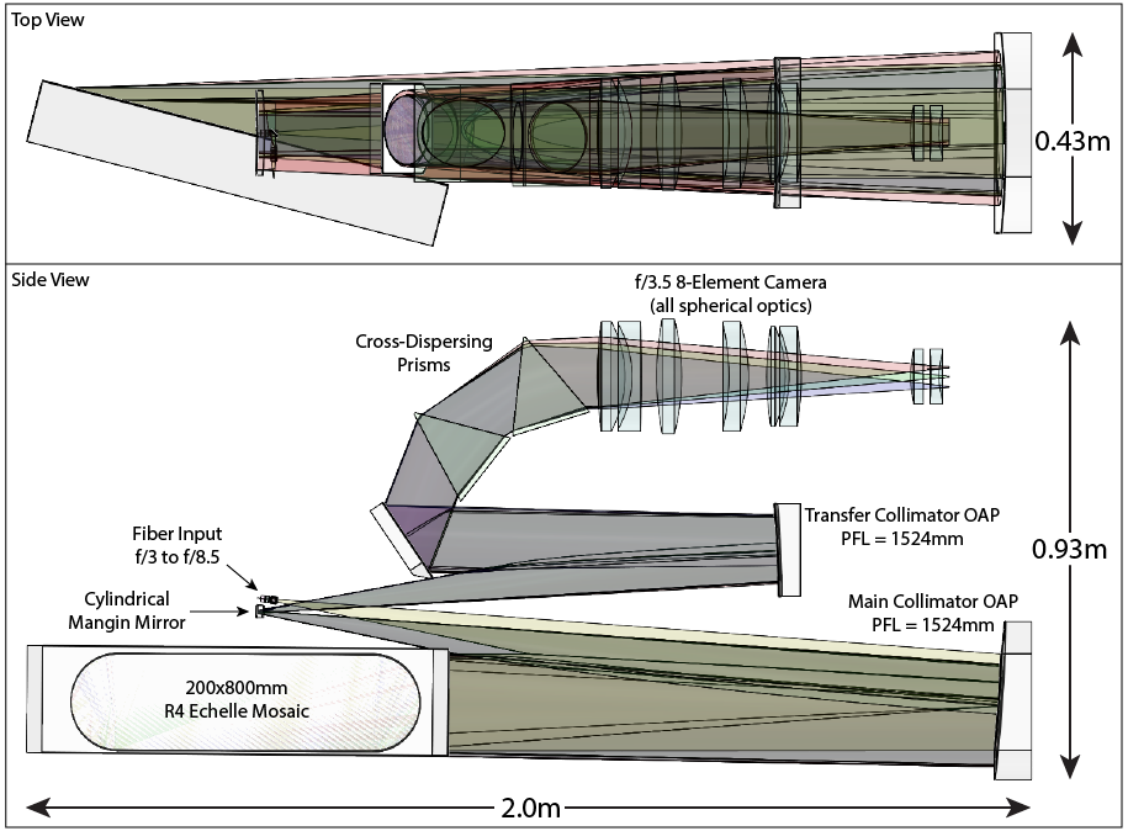

Figure 9. *Top Panel*: Top-down view of the spectrograph optical layout. *Bottom Panel*: Side view of the spectrograph optical layout. The total optical footprint is 2.0m long, by 0.93m tall, by 0.43m wide.

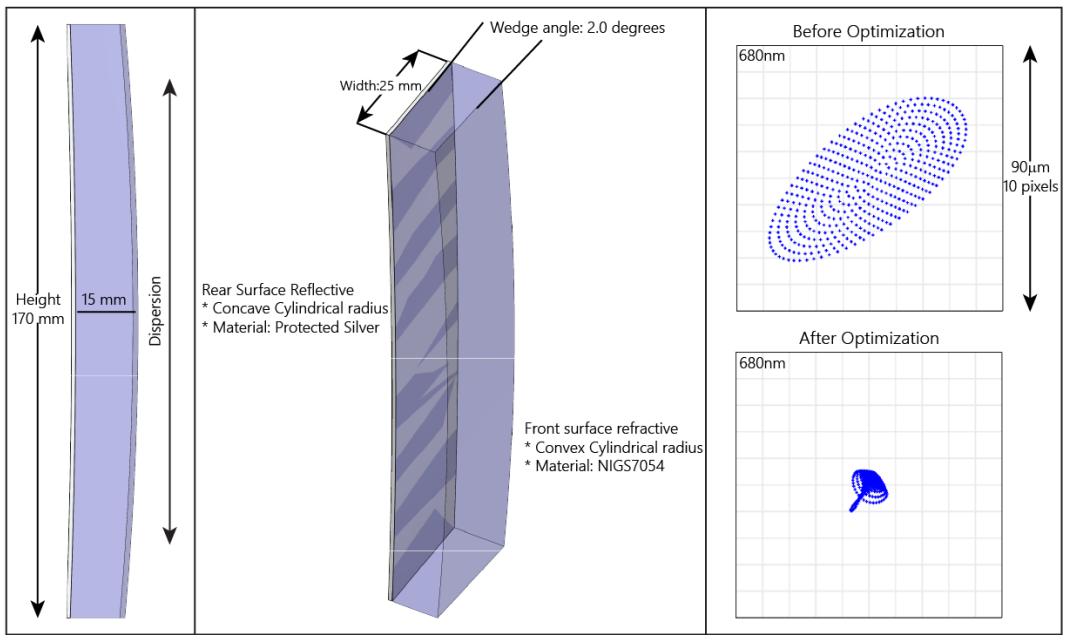

Figure 10. *Left Panel:* Profile view of the Mangin mirror showing the direction of dispersion. *Middle Panel:* Isometric view of the mirror. The front surface is wedged relative to the rear to eliminate ghosting. *Right Panel:* Spot diagrams at 680nm showing the improvement after optimization of the Mangin mirror cylindrical radii.

The front surface of the Mangin mirror is intentionally wedged relative to the rear surface to eliminate ghosting from a first surface reflection. Once the beam has passed through the Mangin mirror, it comes to focus prior to the transfer collimator. The transfer collimator is also an off-axis-parabola with a focal length of 1016mm, yielding an asymmetric coefficient of 0.5. Both the main and transfer collimators are fabricated using off-the-shelf parents from Space Optics Research Labs. This dramatically reduces the fabrication cost from custom OAPs; however it does constrain the optical design space. After the transfer collimator the beam is folded by a flat mirror to bring the optical axis back to being parallel to horizontal following the cross-dispersing prisms. This fold also reduces the overall footprint of the instrument. The prisms are fabricated from Ohara's iLine material PBM2Y, and are polished and coated by the Nikon Corporation.

The camera is f/3.5, composed of eight spherical elements and provides 4-pixel sampling of the resolution element (width of the rectangular fiber). We took a similar approach to the camera optimization that is being employed in other next generation extreme precision RV spectrographs. The basic approach is to build a merit function that provides a highly uniform and simple line spread function (LSF), as well as make it insensitive to a changing pupil illumination[13]. Since perfect scrambling can never be achieved, the latter provides another level of scrambling to help minimize RV errors due a changing shape and position of the LSF.

Spot diagrams for seven wavelengths in seven different orders spanning the bandpass of EXPRES are shown in **Figure 11**. The order number is listed on the left hand side, and the wavelength in Ångstroms is listed above the spot. Each spot is bound by a 4x4 pixel (36μm) box, representative of the width of the 1x4 rectangular input fiber. All of the spots could effectively be bound by a 2x2 pixel box, and are therefore well within the width of the resolution element. They are relatively round and smooth, and uniform across the array. The right panel of **Figure 11** shows color gradient plots of the distribution of the spot sizes on the array for two different optimizations of the camera. The top figure in the right panel illustrates the distribution of spot sizes after an initial optimization using a standard spot minimization merit function. The bottom is the distribution after the optimization described above was employed. These plots contrast a uniform and symmetric camera response against that with an asymmetric, non-uniform response.

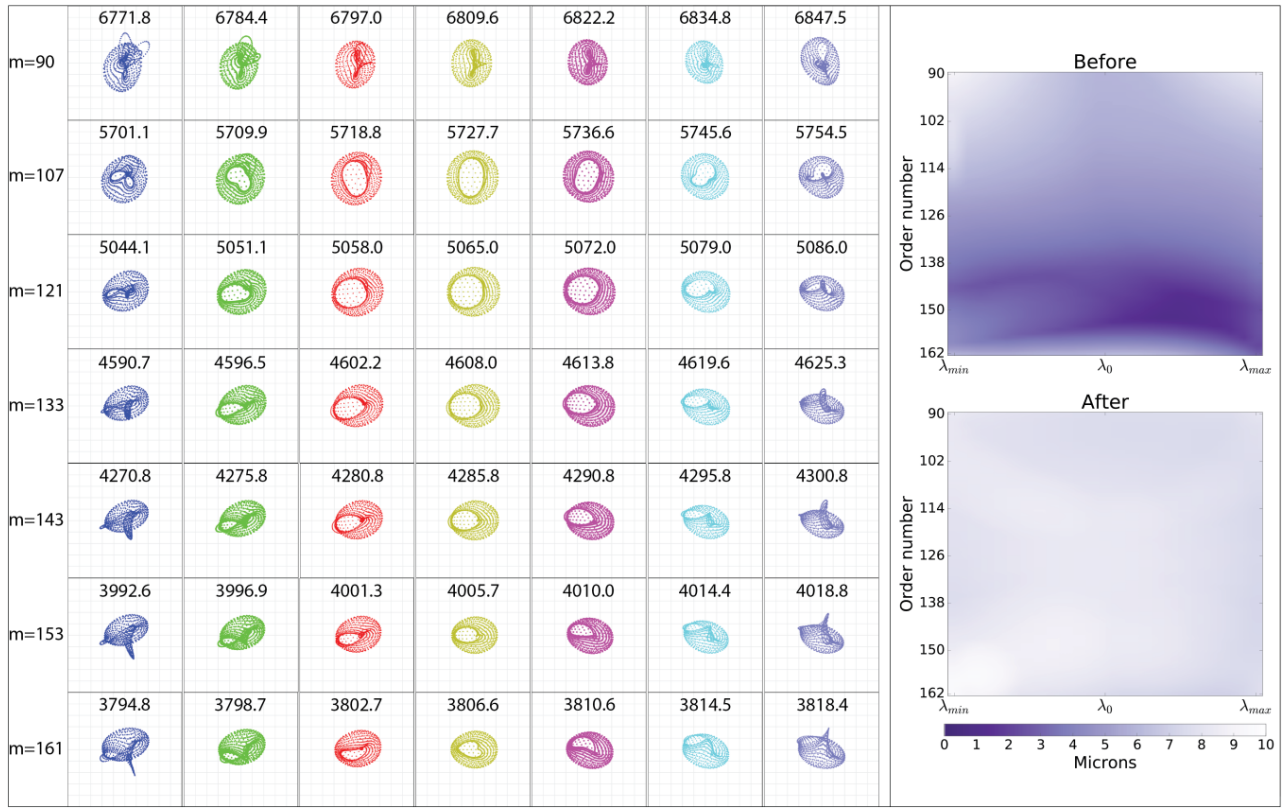

**Figure 11**. *Left Panel*: Spot diagrams for seven wavelengths in seven orders for the entire bandpass of EXPRES. *Right Panel*: Color gradient plots contrasting two different camera optimization routines. The top plot used a standard spot minimization merit function, and the bottom used the routine described in the text, to give a uniform, symmetric response.

## 3.3 Mechanical Design

All optomechanical components for the FEM are either being fabricated, or have been delivered. Laboratory assembly, integration, and test is set to begin in August of 2016. The spectrograph vacuum enclosure is finalized and about to start fabrication. The mechanical design of the optical bench and cryostat is nearing final. All optics have either been received or are in the fabrication process.

### 3.3.1 Front-End Module

**Figure 12** shows the optomechanical design of the FEM. The outer housings are highlighted in the left panel of the figure. The structure is composed of four main bodies. There is an aluminum ring that interfaces to the telescope, and then three Invar bodies that bolt together and house the optics. Calibration light is transported from the spectrograph room via a circular fiber, and is injected into the beam path via a deployable mirror. The right panel is a cut-away view exposing the inside of the FEM. The FEM is environmentally sealed with the exception of a retractable cover that can be remotely actuated to open when the instrument is operational, and then closed when not in use to prevent dust and other contaminants from entering the structure. There is an aperture stop located at the telescope focus that can be exchanged with an alignment target. Alignment fiducials can be placed throughout the optical train that enable the FEM axis to be aligned with the telescope optical axis. Translation and tip-tilt adjustments on both the fold an FTT mirrors allow for adjustment.

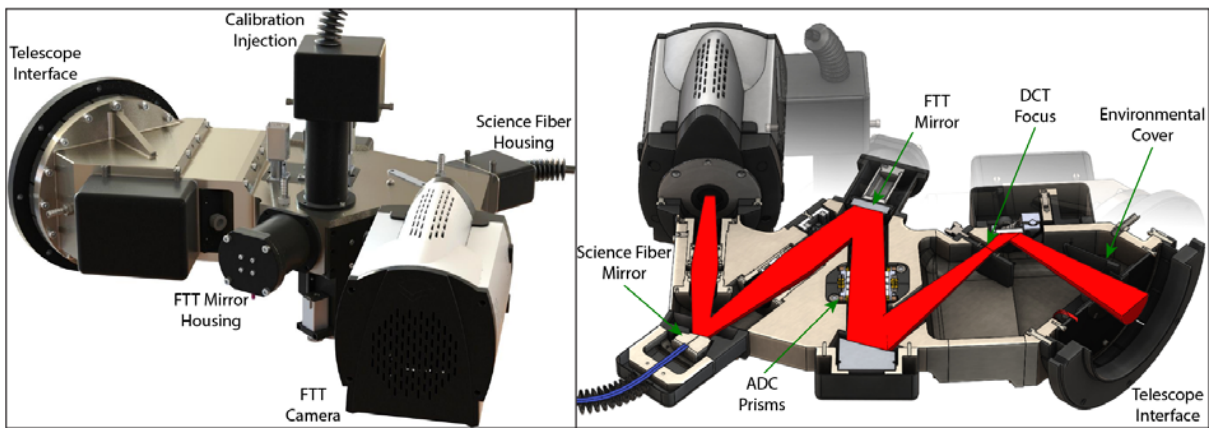

**Figure 12**. *Left Panel*: The outer mechanical housings of the FEM. *Right Panel*: Cut-away of the mechanical housings exposing the opto-mechanisms within.

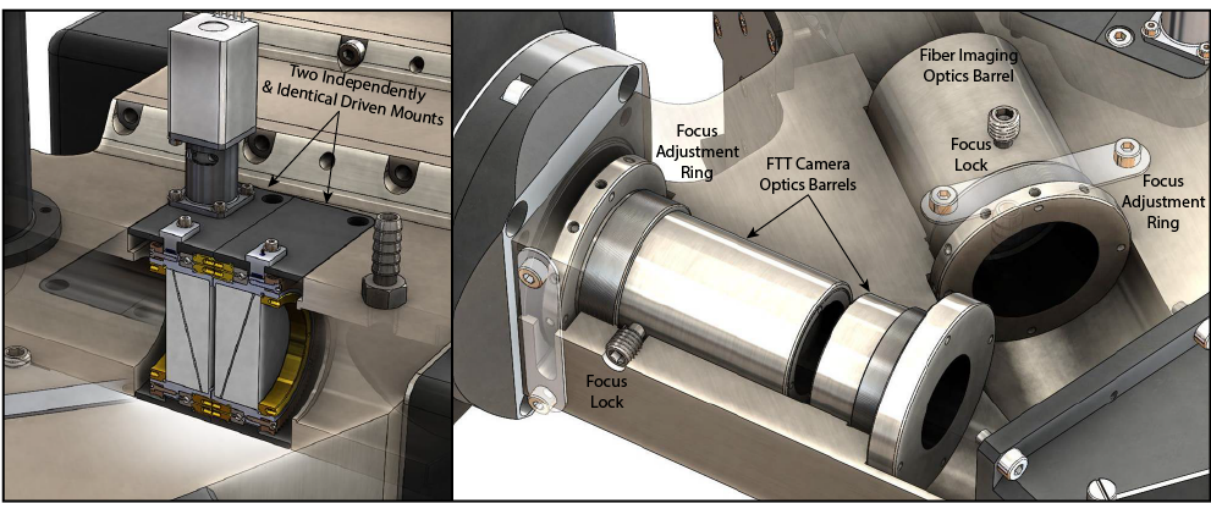

**Figure 13**. *Left Panel*: Cut-away of the FEM highlighting the ADC mechanisms. *Right Panel*: Cut-away of the FEM highlighting the fiber and FTT camera lens barrels.

The left and right panels of **Figure 13** highlight the ADC and lens barrel assemblies. The ADC prism pairs are housed in identical mounts that can be independently driven. The drive mechanism is a stepper motor with a worm/gear combination. It provides for 1.08 arcminute per step precision. Magnets mounted in the rotating gear trip a fixed Hall sensor to locate a home position. From there encoder counts from the home position control the speed and orientation of the prism pairs. Focus onto the fiber mirror and FTT camera is done manually during alignment. The lenses are mounted into barrels that can be translated via rotation and then locked into place with a no-mar set screw. Only the triplet in the FTT camera optics is translatable, as the doublet is fixed. The Invar housing will maintain drifts within the depth of focus of the optics over the annual temperature range of the observatory site. Focus can then be acquired and kept by adjusting the telescope secondary and FTT camera.

### 3.3.2 Spectrograph

Currently the DCT does not have a suitable room to house the EXPRES spectrograph. Part of the development of this instrument is to have this room constructed. It will provide a light-tight, climate controlled laboratory environment with anti-static flooring and a vibrationally isolated slab for the spectrograph vacuum enclosure. The temperature will be controlled to 293 +/-0.5K over a 24 hour period, with a relative humidity of 40 to 60%. The spectrograph room will also house the LFC optics components, slicer/double-scrambler, and calibration assembly. The room will have a dedicated HVAC system, and will be hooked into a back-up power generator in the case of power failure. To reduce the workload on the HVAC system as well as the as the electronic noise in the room, the majority of the electronics power supplies will be housed in racks outside of the room.

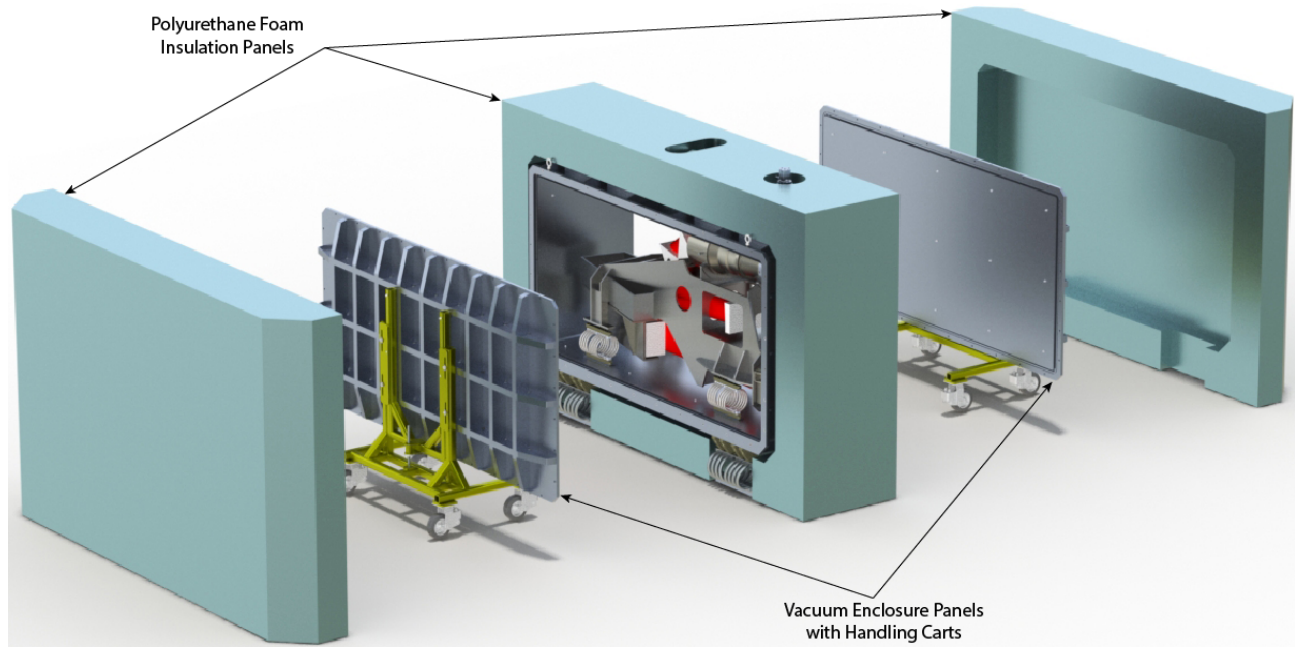

**Figure 14**. An exploded view of the spectrograph thermal and vacuum enclosures, exposing the optical bench.

EXPRES will not employ active temperature control beyond that of the room. Figure 14 is an exploded view of the spectrograph showing the thermal enclosure (polyurethane panels) surrounding the welded aluminum vacuum enclosure, and 304 stainless steel optical bench. The thermal enclosure is not in contact with the vacuum enclosure or the vibrationally isolated slab that it sits upon, thus minimizing heat loads as well as providing an acoustical barrier for vibrations. The left panel of **Figure 15** shows the vacuum enclosure and optical bench. As mentioned in Section 3.3.2, the vacuum enclosure is bolted to a concrete slab in the instrument room that is isolated from the building foundation, and telescope pier. Vibration dampeners then isolate the slab from the enclosure, and the enclosure from the optical bench. To minimize thermal conduction, G10 blocks serve as the mechanical interfaces on either side of the vibration dampeners.

All of the optics are mounted kinematically into cells, which are then mounted kinematically onto the welded optical bench. The top of the vacuum enclosure is divided into five panels that can be custom configured depending upon the feedthrough requirements. There are no optically transmitting windows; therefore a single panel on the top of the enclosure has been dedicated to feed the fibers in and out of the vacuum. A second panel is dedicated to the vacuum pump station, which consists of roughing and turbo pumps, and vacuum gauge. There are two different vacuums being pulled on the instrument. The one just described is for the vacuum enclosure itself which is required to be maintained at $10^{-3}$ Torr. The second vacuum is that for the cryostat, which occupies a third panel on the enclosure. A Stirling cryocooler interfaces to the cryostat to cool the detector. Further vibration dampening is deployed between the cooler and detector cold plate. An active vibration cancellation system minimizes vibrations at the cooler cold tip, and a passive system is employed between the cold tip and detector cold plate to remove any residual vibrations. Sorbothane rings isolate the cryocooler from the vacuum enclosure.

The requirement on the thermal stability of the optical bench is +/-1mK over a 24 hour period. While the spectrograph room has active temperature control, the spectrograph is passively controlled. A thermal study was conducted to verify that the passive system could meet the optical bench temperature requirement. The thermal model assumes both convective and radiative heat transfer paths from the room all the way through to the optical bench. The top plot in the right panel of **Figure 15** shows the response of the thermal enclosure, vacuum enclosure, and optical bench to a room that varies by +/-0.5K over a 24 hour period. The results show an optical bench temperature variation of couple-tenths of a milli-Kelvin. It is also useful to look at the temperature rate of change of the optical bench over this period of time, and compare that to the longest detector exposure of 20 minutes. The bottom plot in the right panel of **Figure 15** shows the temperature rate of change of the optical bench over this same 24 hour period.

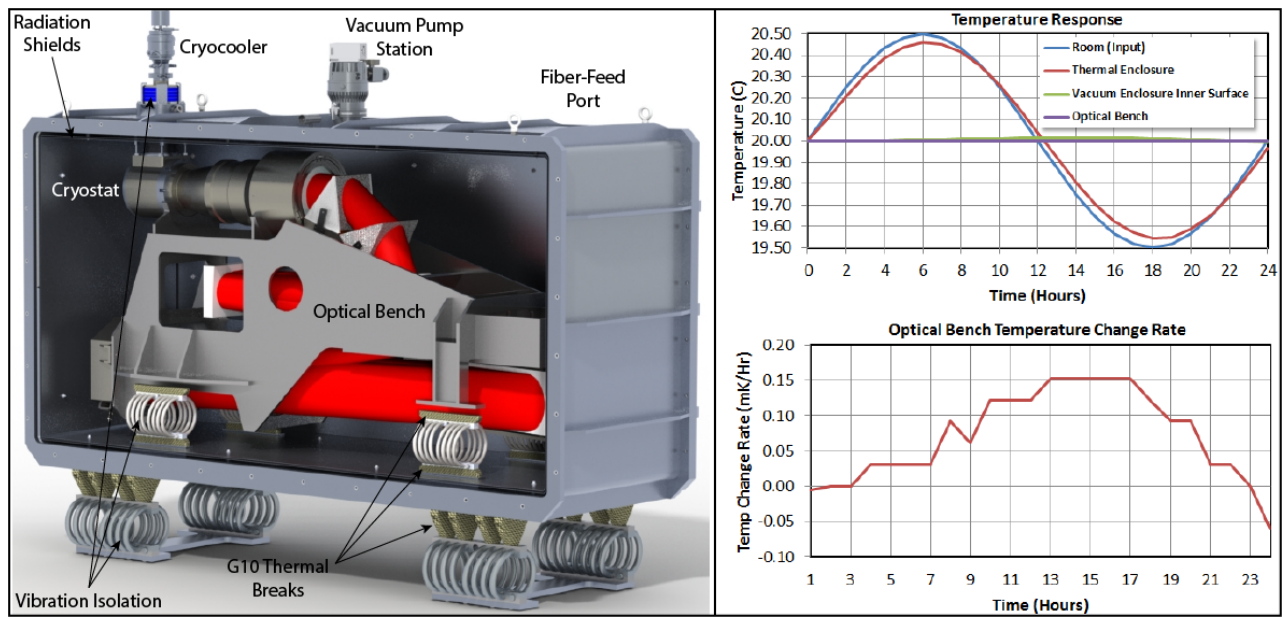

**Figure 15**. *Left Panel*: Spectrograph vacuum enclosure calling out vibration isolation, radiation shields, and vacuum ports. *Right Panel*: Top plot is the thermal response plot of the thermal enclosure, vacuum enclosure, and optical bench in response to the room. Bottom plot is resulting rate of temperature change of the spectrograph optical bench over a 24 hour period.

## 4. STABLE ILLUMINATION

Instrument stability in the quest for pushing the RV precision to the level necessary to detect exo-Earths must be approached on multiple fronts. It is key to provide a stable optical input to the spectrograph in addition to a stable environment. Studies have shown variations in the illumination of the spectrograph can result in shifts on the detector masking sensitive Keplerian motions[14]. EXPRES mitigates the issue in three key areas: by stabilizing the image on the science fiber, by creating a fiber link that decouples the science fiber input from the spectrograph input, and by optimizing the camera optics within the spectrograph to be insensitive to variations in pupil illumination.

## 4.1 Stabilized Light Injection

Modern telescope facilities are designed and operated to minimize the effects of dome seeing but even with such practices, residual effects remain from temperature fluctuations and differences throughout the dome and telescope and winds through the spider support structure of the telescope itself. Coupled with the natural seeing, results in wavefront errors with the single largest component being attributed to wavefront tilt or image motion. Image motion causes the beam to wander on the science fiber decreasing coupling efficiency and resulting in variations in the spectrograph illumination due to imperfect fiber scrambling. It has been suggested that by correcting for image motion, an approximate 15% decrease in the FWHM of image may be seen[15]. As a goal to maximize throughput and instrument efficiency, EXPRES incorporates a Fast Tip-Tilt (**FTT**) system.

Power spectral density measurements of image motion provided by the DCT showed spectral features at discrete frequencies up to approximately 60 Hz. Basic control theory suggests a servo system should operate at 5-10 times the highest correction frequency. Thus, the FTT system was designed to operate at sampling rates greater than 600 Hz so as to sample and correct the highest frequency at least 10 times per cycle (the coherence of these frequencies are not well understood). The FTT system is tasked with stabilizing the image on the science fiber to less than 0.05 arcseconds at magnitudes well beyond $m_v=7$, simulations have demonstrated neither requirement to be an issue. **Figure 16** shows the system architecture.

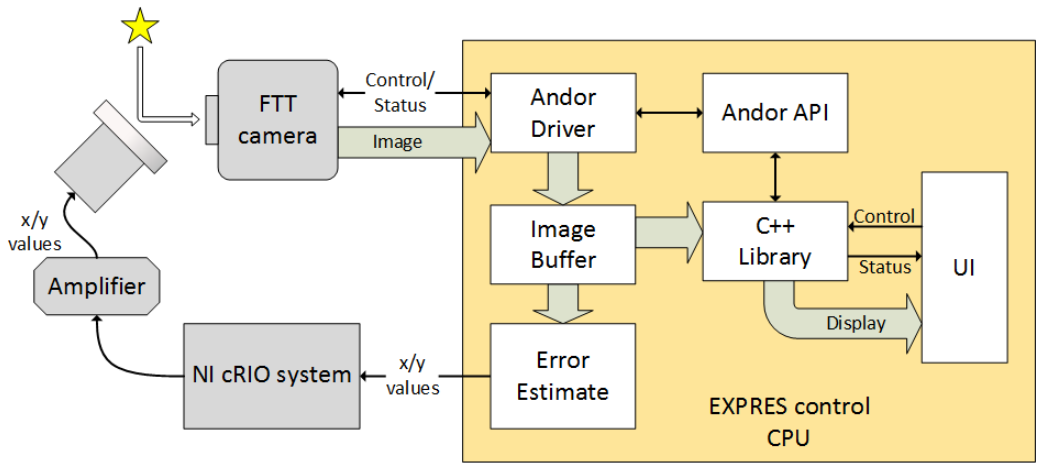

**Figure 16**. Diagram of the EXPRES fast tip-tilt system. The FTT system is to stabilize image motion on the science to less than 0.05 arcseconds at rates exceeding 600Hz. The entire system is comprised of off-the-shelf components and the camera is controlled by a standard rack mount computer.

The FTT system operates in two modes: full frame mode and closed-loop mode. In full frame mode, the 80 arcsecond field is used for target acquisition and identifying the location of the fiber on the camera array. Closed-loop mode reads a specified sub-frame from the camera and provides feedback to the FTT mirror stage. The design was baselined for a 5x5 arcsecond field corresponding to a 32x32 pixel sub-frame. Wavefront tilt is corrected for in the pupil plane by reimaging the DCT primary onto the FTT mirror, a 38.1 mm mirror mounted on a Physiks Instrumente stage. The resulting angular magnification is 168 meaning a 0.05 arcsecond wavefront tilt error on the primary is corrected for with a 4.2 arcsecond tilt of the FTT mirror. After correction, light is directed through the fiber imaging triplet and coupled into the science fiber. Spillover light not coupled is reflected from the fiber embedded mirror reimaged onto the FTT camera, an Andor iXon Ultra 897, at a 0.16 arcsecond per pixel plate scale. A rack mounted computer in the observatory server room controls the camera and reads the data over a fiber based USB extender. With the servo enabled, error estimates are fed back to the mirror through a drive circuit comprised of a National Instruments compact RIO system and PiezoDrive amplifier. A custom software library was designed and implemented for straightforward integration into the EXPRES control software. It provides a high level set of system calls and a copy of the image buffer for display, logging, and diagnostic purposes. All components within the system are off-the-shelf. Prototype testing has demonstrated as system latency of less than 0.5ms from the end of an integration time to amplifier output. This latency includes the camera frame transfer, subsequent readout, error estimate, and control system update.

## 4.2 Pupil Slicer & Double-Scrambler

EXPRES is a fiber-fed instrument so as to locate the spectrograph in a stabilized environment. An additional benefit is that fibers have an inherent scrambling property that decouples the output illumination from variability at the input[16]. The level of scrambling may be enhanced with non-circular fibers[17,18] and the use of a double scrambler that exchanges the near field and far field between a pair of fibers[19]. EXPRES uses both techniques: an octagonal science fiber feeds a double scrambler that outputs to a rectangular fiber coupled to the spectrograph. Within the double scrambler is a Bowen-Walraven type pupil slicer that slices and stacks the beam to boost the resolution by, in effect, 'narrowing the slit' without apodizing the beam.

The combined pupil slicer and double scrambler design and fabrication was contracted to Fraunhofer IOF in order to leverage their expertise and precision machining technologies including alignment turning of mounted lenses, diamond machining of optical and reference surfaces, and precision machining of other components critical to alignment. A ray trace of the design is shown in **Figure 17**. The design is anticipated to have a throughput of better than 95% averaged over the total bandpass. In the design, the 66 um octagonal fiber coming from the FEM (d1) plugs into the unit. L1 images the pupil onto the Bowen-Walraven slicer comprised of a pair of mirrors. L2 and L3 reimage the sliced pupil onto the 33 µm x 132 µm rectangular fiber (d2) that feeds the spectrograph. Both fibers are connectorized with Diamond Mini-AVIM connectors for superior stability and alignment repeatability.

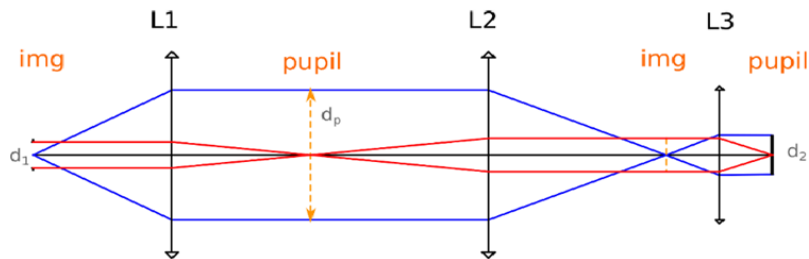

**Figure 17**. A ray trace of the combined pupil slicer and double-scrambler integrated into the EXPRES fiber link. The octagonal fiber enters from the left (d1). The input is sliced once in the pupil plane (dp) with the far-field of the octagonal fiber being projected onto the near-field of the rectangular fiber exiting on the right (d2).

Aspects critical to throughput and performance of the DSC identified by Fraunhofer include alignment of the input fiber (d1) to the optical axis as it is a telecentric system, the combined tilt of the input fiber and first lens (d1 and L1) relative to the pupil slicer, and the parallelicity of the slicer mirrors themselves as misalignment will affect one half of the pupil injected into the rectangular fiber. Specifications from Fraunhofer state alignment of the DSC will include centration and tilt of the input optics (d1 and L1) to <1 µm and <<1 millradian respectively, parallel mirror slices, and centration of the output optics (L3 and d2) to <1 µm. Alignment will also include proper rotation between the flats of the octagonal input to the flats of the rectangular output once propagated through the system. A rendering of the design is shown **Figure 18**. To minimize moving parts in the spectrograph and locate parts prone to failure outside of the spectrograph, the shutter was located in the double scrambler. As shown in **Figure 18**, a solenoid based rotary shutter is inserted into the unit. The shutter is vibrationally isolated from the assembly and everything is enclosed within a light-tight box.

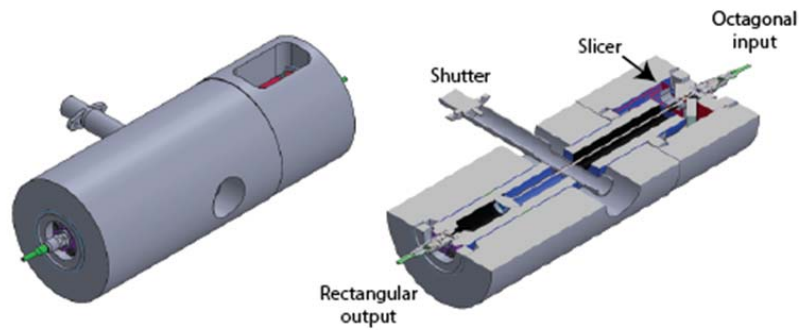

**Figure 18**. Rendering of the combined pupil slicer and double-scrambler. For scale, the unit is 155mm in length and 60mm in diameter. Note a solenoid based cylindrical shutter mounted has been integrated into the assembly.

# 5. PRECISE CALIBRATION

Calibration of RV spectra to a precision necessary to detect exo-Earths requires careful consideration of the calibration system. The standard observation mode calls for science exposures to be book-ended with wavelength calibration data. This necessitates the need for a highly efficient and robust system. For the EXPRES calibration system, a high priority was placed on minimizing failure modes, providing an even illumination across the bandpass, making use of extremely accurate and stable light sources, and coupling them efficiently into the spectrograph. The calibration unit architecture is shown in **Figure 19**. Multiple sources can be fed into the calibration fiber (F3) that runs up to the FEM where it is injected into the science path with a retractable fold mirror and ultimately into the science fiber (F4) at the same focal ratio. Calibration sources injected in this manner traverse the same path through the instrument as science targets and produce calibration spectra on identical pixels of the detector. Two additional lights paths are provided: one for simultaneous wavelength calibration to be taken with science observations (fiber F5) and another to produce an "extended" flat-field spectrum that overfills the science orders (fiber F6).

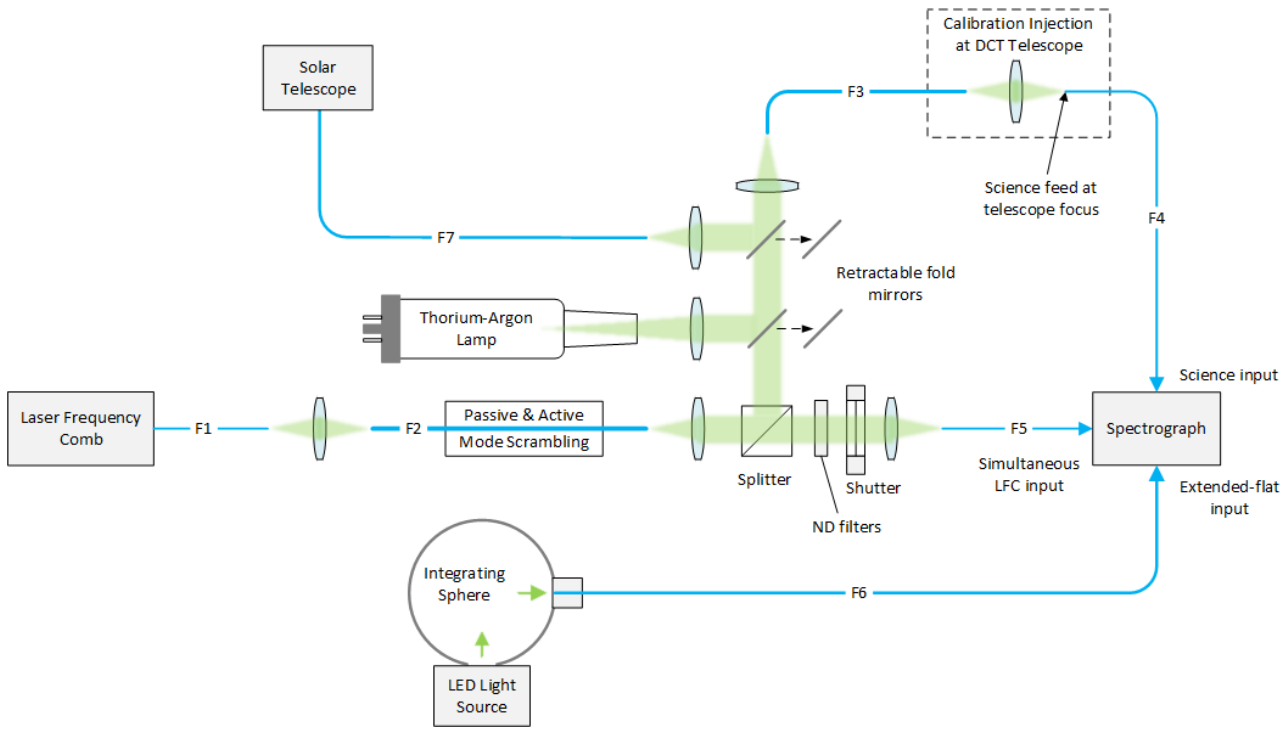

Figure 19. EXPRES calibration unit architecture. Calibration light from a laser frequency comb, LED white light source, Thorium-Argon hollow cathode lamp, and solar telescope is coupled to the spectrograph through a combination of optics and fibers.

## 5.1 Wavelength calibration

A Menlo Systems laser frequency comb (LFC) will be the primary wavelength calibrator for EXPRES. The LFC was selected for the following reasons:

- It covers a wavelength range of 450-700 nm, which is well matched to our instrument bandpass,
- It produces highly stable, evenly spaced lines at 14 GHz (about 10 pixels) across the full wavelength range,
- The photon luminosity of the lines can be tuned for an inverse response of the instrument, which allows us to produce a uniform intensity spectrum on the CCD,
- It is available as a turnkey system with support from Menlo systems,
- It has demonstrated performance on other high precision RV spectrographs.

EXPRES will provide two modes for wavelength calibration using the LFC. The standard operational mode will be to bracket each science exposure with an LFC spectrum injected through the science fiber. This mode will provide the best precision for wavelength calibration as the LFC spectra will sample the same instrument path as the science spectra and be subject to the same instrumental effects such as FRD from fiber bending in the telescope, or fixed pattern noise on the CCD. The standard mode requires only a fold mirror in the FEM to be translated into place. The second mode is a simultaneous comparison mode where the LFC spectrum is obtained along with the science spectrum with the individual orders falling next to each other. The simultaneous mode will be used to monitor and characterize the stability of the instrument during long exposures and provide us a diagnostic tool. Once the instrument is well characterized it is not expected that the simultaneous mode will be used on a regular basis.

The coupling of the LFC to the spectrograph is illustrated in Figure **19**. The LFC is output through a single mode fiber (F1) that we couple to a multi-mode fiber (F2). Since the LFC lines are highly coherent they create interferences in the multi-mode fiber that appear as speckles (modal noise) in the illumination that change with time depending on conditions (e.g. temperature, fiber bending). To mitigate the modal noise, we use a large, circular ~1 mm diameter multi-mode fiber with high NA to support a high number of modes that enable efficient mode scrambling through passive sinusoidal serpentine bends[20] and active mechanical agitation[21] of the fiber. Following the modal noise scrambling the LFC light is split for the two modes – standard and simultaneous. For the standard mode the light is reimaged onto the 200 µm fiber (F3) that feeds up to the FEM for injection into the 66 µm octagonal science fiber (F4). For the simultaneous mode the light is reimaged onto a 33 µm square fiber (F5) that is injected directly into the spectrograph adjacent to the science fiber. The light for the simultaneous mode passes through a neutral density filter wheel to allow the intensity of the simultaneous spectra to be adjusted to roughly match to the intensity of the science spectra (target magnitude), and a shutter to disable the simultaneous mode while the LFC is left on for stability during standard mode observations.

A Thorium-Argon hollow cathode lamp is included in the EXPRES calibration unit as secondary wavelength reference and as well as a backup for the LFC (e.g. during periodic maintenance shutdowns). The ThAr calibration mode is selected by inserting a fold mirror in front of the LFC light path and reimaging the ThAr filament onto the 200 µm fiber. ThAr calibration is not an essential mode of the instrument, thus its fold mirror will be a fail-open type component to prevent the LFC light from being blocked in the case of a failure.

### 5.2 Flat-Fielding

The EXPRES calibration system incorporates novel concepts for improving the flat fielding performance; flat-field spectra that extend beyond the edges of the science orders to allow for high SNR flat-field correction to the edge of the science orders, and an LED white light source that can be tuned to the inverse response of the instrument to produce a flat response on the CCD across the full instrument bandpass.

The "extended" flat spectra are obtained by feeding the white light source through a 60 x 180 µm rectangular fiber (F6) and injecting it into the spectrograph using a pickoff mirror in front of the science fiber. This has the effect of a longer slit which produces a flat-field spectrum with orders that are wider than the science orders. The extended flat fiber includes an oversized simultaneous fiber as well to allow both the science and simultaneous spectra to be corrected using the same flat-field exposures. The extended flats bypass the science fiber, but the errors that are corrected using flat-fields (e.g. CCD fixed pattern noise, ghosting) originate in the spectrograph and have negligible influence from the fibers. In cases where it may be desirable to obtain flat-fields using the science fiber, the DCT dome screen can be used for quartz lamp exposures.

### 5.3 White Light Source

We have been working with FiberTech Optica to design and fabricate an LED light source for the EXPRES calibration system that produces a flat-field spectrum with a signal level of 30,000 e-/pix that is uniform to within 10% across the entire 380-680 nm bandpass in a 10-second exposure. The results of simulations that FiberTech Optica conducted using 18 LED emitters show promise that the power distribution and flatness requirements for EXPRES can be achieved (**Figure 20**). For the proposed design we could increase the number of emitters to 25. The LEDs mount to a cold finger inside an integrating sphere with a thermal electric cooler to stabilize the output. The integrating sphere provides a homogeneous and isotropic illumination of the extended flat fibers. Tests have been conducted at FiberTech Optica to demonstrate that the power level needed to meet the EXPRES requirements can be easily achieved using the integrating sphere to couple the light.

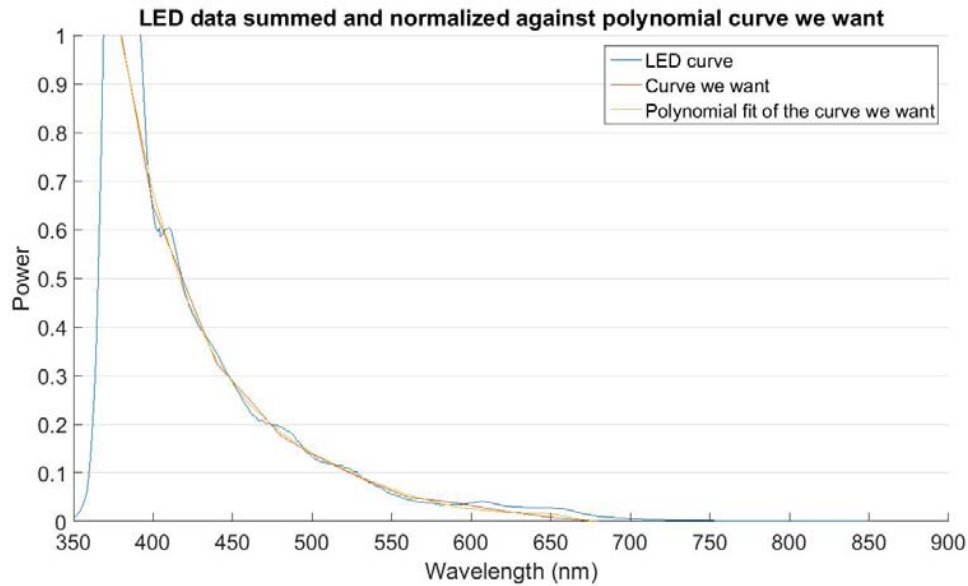

Figure 20. Plot showing the simulated fit of a LED light source (blue line) against the power requirement to produce a flat response on the EXPRES CCD detector (orange line). Plot courtesy of FiberTech Optica.

**5.4 Solar Port**

To efficiently obtain solar spectra on a daily basis for detecting other solar system bodies, the calibration system includes an additional port to the calibration unit via fiber (F7). This port is selected by inserting a fold mirror in front of the LFC light path and reimaging the fiber (F7) onto the calibration fiber. This is not an essential mode of the instrument and thus its fold mirror will be a fail-open type component to prevent the LFC light from being blocked in the case of a failure.

## 6. ACKNOWLEDGMENTS

We would like to acknowledge NSF Major Research Instrumentation Award AST 1429365, as well as an NSF Advanced Technologies Instrumentation Award AST 1509436. The author would also like to acknowledge Gabor Furesz for helpful discussions on spectrograph camera optic optimization, and Andrew Szentgyorgyi for valuable input on design considerations.